\documentclass{pasj01}
\begin{document}
\SetRunningHead{K. Sadakane et al.}{Emission lines in  $\iota$  Her}
\Received{}
\Accepted{}

\title{A Spectroscopic Study of Weak Metallic Emission Lines in a B3V Star $\iota$  Her } 

\author{Kozo \textsc{Sadakane},\altaffilmark{1}
         and
        Masayoshi \textsc{Nishimura}\altaffilmark{2}}

\altaffiltext{1}{Astronomical Institute, Osaka Kyoiku University, Asahigaoka,  Kashiwara-shi, 
            Osaka   582-8582}
\email{sadakane@cc.osaka-kyoiku.ac.jp}

\altaffiltext{2}{2-6, Nishiyama-Maruo, Yawata-shi, Kyoto  614-8353 }
%% `\KeyWords{}' always has to be placed before `\maketitle'.
%Do NOT move this preamble from here! 

\maketitle

\begin{abstract}

We present a list of weak emission lines (WELs) observed in a sharp-lined B3 V 
star  $\iota$  Her (HD 160762) using high resolution ($\it R$ = 65000) and high 
SN ($\sim$ 1300) spectral data. The list covers a spectral region between 4900 
\AA~ and 10000 \AA. We register 207 WELs in this star and identified 190 lines
including ten ions (nine elements). Emission lines of C~{\sc ii}, N~{\sc i}, 
 Cr~{\sc ii}, Mn~{\sc ii},  and Ni~{\sc ii} have been identified among 
normal B-type stars for the first time. 17 emission lines remain unidentified.
%Observed wavelengths, emission peak intensities, emission equivalent widths
%and relevant atomic data  are given for identified lines. 
We compare our list with the published list of WELs for 3 Cen A  
\citep{wahlgren2004}  and found that numbers of detected emission lines reflect 
differences  in abundance between these two stars. We detect 13 C~{\sc i} emission
lines in $\iota$  Her (normal  in C), while only one C~{\sc i} emission line
is found in 3 Cen A (deficient in C). Many emission lines of P~{\sc ii} and Cu~{\sc ii} 
have been detected in 3 Cen A (overabundant in both P and Cu), while no
emission line of these ions has been found in  $\iota$  Her. 
Many emission lines of Fe~{\sc ii} are visible in the shorter wavelength side of  6000 \AA~ in  
$\iota$  Her, while these emission lines  are missing in 3 Cen A. Close inspections 
of spectral data of 3 Cen A reveal that apparently missing  Fe~{\sc ii} lines appear
as absorption lines in this star. 
Because these two stars have nearly the same atmospheric parameters 
 ({\it $T_{\rm eff}$} and log $\it g$), a physical interpretation which is independent on
these two parameters  is needed to account for this observation.

\end{abstract}

\section{Introduction}

The presence of sharp and weak emission lines (WELs) in optical
region spectra of B-type stars has been initially reported by \citet{sigut2000}
in a  ${}^{3}$He star 3 Cen A (HD 120709). They found  sharp emission 
lines of  Mn~{\sc ii}, P~{\sc ii}, and Hg~{\sc ii} ions in the red spectral region.  
They also found very weak emission lines of Mn~{\sc ii}  in a mild HgMn star 
46 Aql (HD 186122, B9 III).
\citet{wahlgren2000} reported detections of weak emission lines 
originating from high excitation states of Ti~{\sc ii}, Cr~{\sc ii},  and Mn~{\sc ii} in 
 late B-type HgMn stars. They showed that emission lines of  Mn~{\sc ii} are
absent in HgMn stars for which the Mn enhancement is greater than 1.3 dex
and suggested a dependence of strengths of emission lines on the element
abundance.
\citet{wahlgren2004} published a list of nearly 350 emission lines of 3 Cen A
observed in optical and near-IR spectral regions and identified many emission 
lines of P~{\sc ii}, Mn~{\sc ii}, Fe~{\sc ii}, Ni~{\sc ii},  and Cu~{\sc ii}.      
Since then, WELs have been reported mainly among B-type chemically
peculiar stars. 
%HD 175640, HR 6000, a Cen, HD71066, HD19400
\citet{castelli2004} reported many emission lines  of Ti~{\sc ii} and Cr~{\sc ii}
and two emission lines of C~{\sc i} at 8335.15 \AA~ and 9405.72 \AA~ in a HgMn star
HD 175640 (B9 V). According to their analysis covering  $\lambda$$\lambda$
from 3040 \AA~ to 10000 \AA, emission lines are selectively found  for
high excitation lines having large transition probabilities 
(log $\it gf$  $ \textgreater $ --1.0) and  these emission lines are found 
in the red part ($\lambda$  $ \textgreater $  5850 \AA) of the spectrum.
\citet{castelli2007} listed WELs of Cr~{\sc ii}, Mn~{\sc ii}, and
Fe~{\sc ii} in a Bp star HR 6000 (HD 144667).  Time variable emission lines of Si~{\sc ii},
Mn~{\sc ii}, and Fe~{\sc ii} were reported in a magnetic Bp star a Cen (HD 125823)
by \citet{hubrig2007}.  Weak emission components of C~{\sc i}, Ti~{\sc ii}, Cr~{\sc ii}, 
and Mn~{\sc ii} in a HgMn star HD 71066 and  those of
multiplet 13 lines of Mn~{\sc ii} in a PGa star HD 19400 were suggested by
\citet{yuce2011}  and by \citet{hubrig2014}, respectively. 

\citet{sadakane2017} presented observations of WELs of Si~{\sc ii} and Al~{\sc ii}
near 6240 \AA~ in many early B-type stars. They demonstrated that the WELs 
phenomenon is not an exceptional case observed only among chemically peculiar 
stars but generally observed in many $\it normal$ stars.  

Sigut (2001a, b) analysed WELs of Mn~{\sc ii}  
observed in 3 Cen A and 46 Aql and showed that these emission lines can be 
 explained by interlocked nonlocal thermodynamic equilibrium (NLTE) 
 effects combined with the presence of vertical 
stratification. Alternatively, 
Wahlgren and Hubrig (2000, 2004) discussed a mechanism that the 
population of highly excited states might be caused by excitation from the
far-UV continuum radiation in hot stars.
 \citet{wahlgren2008} noted that NLTE effect can be a significant
mechanism for the production of WELs. \citet{sigut1996} carried out 
NLTE radiative transfer calculations  
of near IR region WELs of Mg~{\sc ii} ion in A and B-type stars and published 
predictions of line profiles of four  Mg~{\sc ii} lines between 1.01 $\mu$m and 
4.76  $\mu$m. 

Recently, \citet{alexeeva2016} and \citet{sitnova2018a} constructed 
 comprehensive model atoms for C~{\sc i} and C~{\sc ii} and for Ca~{\sc i} 
and Ca~{\sc ii}, respectively,  and computed the NLTE  line 
formation for these ions. \citet{alexeeva2016} analysed profiles  of four 
C~{\sc i} lines (8335.15 \AA, 9078.28 \AA, 9088.51 \AA~ and 9405.72 \AA)
and two C~{\sc ii} lines (3918.96 \AA~ and 5143.49 \AA)  observed in  $\iota$ Her 
and found that the  emission lines C~{\sc i} and absorption lines of C~{\sc ii}
were finely reproduced by their NLTE calculations. 
\citet{sitnova2018a} demonstrated that profiles of four emission lines of Ca~{\sc ii}
(6456.88 \AA, 8912.07 \AA, 8927.36 \AA~ and 9890.63 \AA) observed in  $\iota$ Her 
could well be reproduced by their NLTE calculations. 
\citet{sitnova2018b} found some emission lines of Fe~{\sc ii}  in the near IR
spectrum of $\iota$ Her. They discussed that over-ionization of Fe~{\sc ii}  ion
will result in weakened Fe~{\sc ii} lines and noted that these lines will either
disappear or come into emission in case of highly excited lines. 

\citet{sitnova2016} constructed a model atom for Ti~{\sc i} ~ and Ti~{\sc ii} and
performed  NLTE line formation calculations for these ions. Their target stars
are A to K type stars including metal-poor stars  with [Fe/H] down to --2.6 dex.       
\citet{alexeeva2018} carried out NLTE line formation calculations for Mg~{\sc i}  
and Mg~{\sc ii} lines for B-A-F-G-K stars. They predicted emission lines of Mg~{\sc i}  
in the infra-red region for F-G-K stars.

Since the pioneering work of \citet{wahlgren2004} on the  ${}^{3}$He star 3 Cen A,  
no list of WELs in early B-type star has been published. The phenomenon of WELs
is observed not only in chemically stars but also among normal early type 
stars. Thus, comprehensive lists of emission lines for representative stars of
several spectral types are needed in order to improve the theoretical  understanding of
the formation mechanism of emission lines.  Our primary goal of
the present study is to provide a comprehensive list of WELs of as many elements  
for a bright and sharp-lined  normal  B3 V star $\iota$ Her based on high quality 
observational data.  We will compare our list with that of 3 Cen A 
\citep{wahlgren2004} and discuss some differences between these two stars. 
We expect the list to be used as a reference guide and hope
to be compared with results of theoretical studies in the future.

\section{Observational data}

Spectral data of four stars ($\iota$ Her,  3 Cen A,   $\gamma$ Peg (HD 886), and 21 Peg 
 (HD 209459))  in the visible to near IR spectral ranges  were obtained with the 
Echelle Spectro Polarimetric Device for the Observation of Stars (ESPaDOnS)\footnote{http://www.cfht.hawaii.edu/Instruments/Spectroscopy/Espadons/} 
attached to the 3.6 m telescope of the Canada-France-Hawaii 
Telescope (CFHT) observatory located on the summit of Mauna Kea, Hawaii. 
Observations with this spectrograph cover the region from 3690 to 10480 \AA, 
and we use data from 3855 \AA~ to 9980 \AA~ in the present study. 
Calibrated intensity spectral data were extracted 
from the ESPaDOnS archive through Canadian Astronomical Data Centre (CADC). 
The resolving power is  $\it R$ = 65000.  Details of observational data used 
in the present study are summarized in table 1.
For $\iota$ Her, we use data observed on two nights (2010 July  20 and 2012 
June  25). 
After averaging downloaded spectral data of each star, we converted the wavelength scale 
of  spectral data of each star into the laboratory scale using measured wavelengths
of five He~{\sc i} lines (4471.48 \AA, 4713.15 \AA, 4921.93 \AA, 5015.68 \AA,
and 5875.62 \AA). Errors in the wavelength measurements are around $\pm$ 3 
km s${}^{-1}$ or smaller. 
 Re-fittings of the continuum level of each spectral order were carried out 
using  polynomial functions. The signal-to-noise ratios (SN) have been 
measured at 13 points from 3900 \AA~ to 9900 \AA, using several line-free
windows within $\pm$ 20 \AA~ of  the specified wavelength and results are 
given in table 2. In the red and near IR spectral regions, SN ratios are somewhat
 uncertain because of
crowding telluric absorption lines. We find that SN ratios of our data become
the peak in the blue-visible region from $\sim$ 4500 \AA~ to $\sim$ 5500 \AA.

%******************************
%*********************************************
\setcounter {table} {0}\begin{table}
      \caption{Summary of observational data} \label{first}
%\tiny
%\footnotesize
\scriptsize
      \begin{center}
      \begin{tabular}{cccc}
\hline\hline
 Object  & Obs date & Exposure   & Number of images \\
\         &          &   (sec)     &                 \\
\hline
$\iota$ Her & 2010 July 20 &  75     & 20              \\
            & 2012 June 25 &  60     & 50              \\
3 Cen A   & 2005 May 19 &   300     & 2               \\
$\gamma$ Peg & 2008 December 17 & 20     & 4        \\
21 Peg  & 2013 August 17          & 290    & 4          \\
           & 2013 August 23          & 290    & 4         \\
\hline
        \end{tabular}
 \end{center}
\end{table}
%*********************************************
\setcounter {table} {1}\begin{table}
      \caption{SN ratio of used data} \label{second}
%\tiny
%\footnotesize
\scriptsize
      \begin{center}
      \begin{tabular}{cccccc}
\hline\hline
Wavelength & $\iota$ Her$^{A}$ & $\iota$ Her$^{B}$  & 3 Cen A & $\gamma$ Peg  & 21 Peg \\
 (\AA)    &   &   &  & &   \\
\hline
 3900     & 400     & 380    &   230   &   330   & 440    \\
 4500     & 1300    & 1050    &   550 &   800   & 980      \\
 5000     & 1200    & 1100    &   770 &   930   & 1150      \\
 5500     & 1500    & 1250    &   700 &   1100  & 1320     \\
 6000     & 850     & 750    &   700   &   680   & 930      \\
 6500     & 820     & 880    &   540   &   710   & 950     \\
 7000     & 650     & 780    &   510   &   570   & 1050     \\
 7500     & 680     & 640    &   710   &   670   & 1130     \\
 8000     & 700     & 500    &   530   &   680   & 920      \\
 8500     & 630     & 900    &   270   &   480   & 700      \\
 9000     & 550     & 610    &   380   &   550   & 630      \\
 9500     & 510     & 820    &   250   &   350   & 600      \\
 9900     & 380     & 720    &   145   &   220   & 470      \\
\hline
        \end{tabular}

 \end{center}
 A: 2010  July  20, B: 2012 June 25
\end{table}
%****************************************
%***************** figures ***************************
%figure1
\begin{figure}
     \begin{center}
       \FigureFile(90mm,100mm){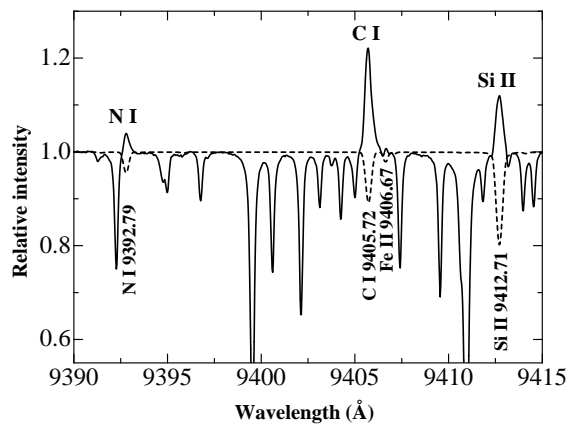}
       %%% \FigureFile(width,height){filename}
     \end{center}
     \caption{A Sample spectrum of   $\iota$  Her near 9400 \AA.
         In addition to the strong emission line of C~{\sc i} at 9405.72 \AA, 
     emission  lines of N~{\sc i} at 9392.79 \AA~ and Si~{\sc ii} at 9412.7 \AA~
     can be confirmed. The Si~{\sc ii} line is actually a blend of two 
     Si~{\sc ii} lines    at 9412.67 \AA~ and 9412.78 \AA. 
     Solid  and dashed lines show observed and simulated spectra, respectively.
   %  The solar abundances have been assumed in the simulation.
   }

    \end{figure}
%----------------------------------------------------
%figure2
\begin{figure}
     \begin{center}
       \FigureFile(90mm,100mm){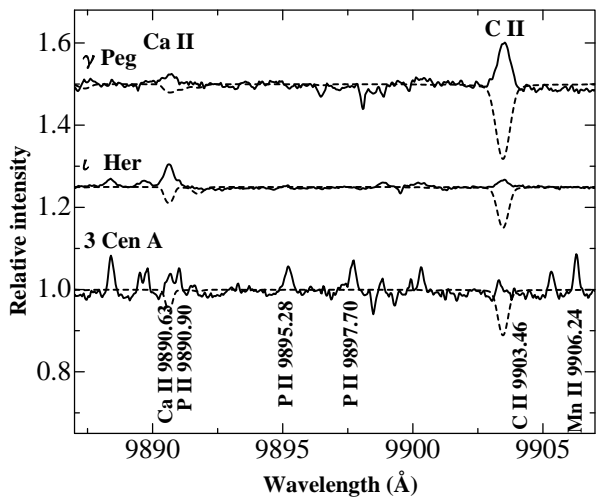}
       %%% \FigureFile(width,height){filename}
     \end{center}
     \caption{A Sample spectrum of   $\iota$  Her near 9900 \AA.
   Data of $\gamma$ Peg and 3 Cen A are shown for comparison. 
Emission lines of Ca~{\sc ii} at 9890.63 \AA~ and C~{\sc ii} 9903.46
are visible in two normal stars.
  Solid  and dashed lines show observed and simulated spectra, respectively. 
 %The solar abundances have been assumed in simulations.
}

    \end{figure}
%------------------------------------------------
%figure3
\begin{figure}
     \begin{center}
       \FigureFile(90mm,100mm){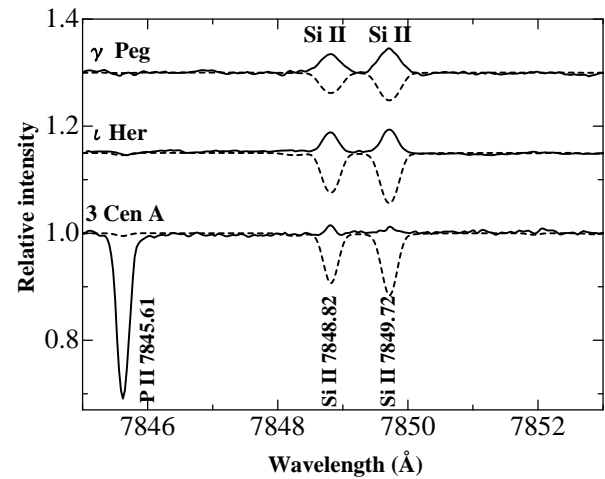}
       %%% \FigureFile(width,height){filename}
     \end{center}
     \caption{A Sample spectrum of   $\iota$  Her near a pair of Si~{\sc ii}
emission lines at 7848.82 \AA~ and at 7849.72 \AA.
   Data of $\gamma$ Peg and 3 Cen A are shown for comparison. 
The weakness of these emission lines in 3 Cen A is to be noted, in which 
the abundance of Si is -0.14 dex lower than that in  $\iota$  Her.
  Solid  and dashed lines show observed and simulated spectra, respectively. 
% The solar abundances have been assumed in simulations.
}
    \end{figure}
%----------------------------------------------------
%figure4
\begin{figure}
     \begin{center}
       \FigureFile(95mm,110mm){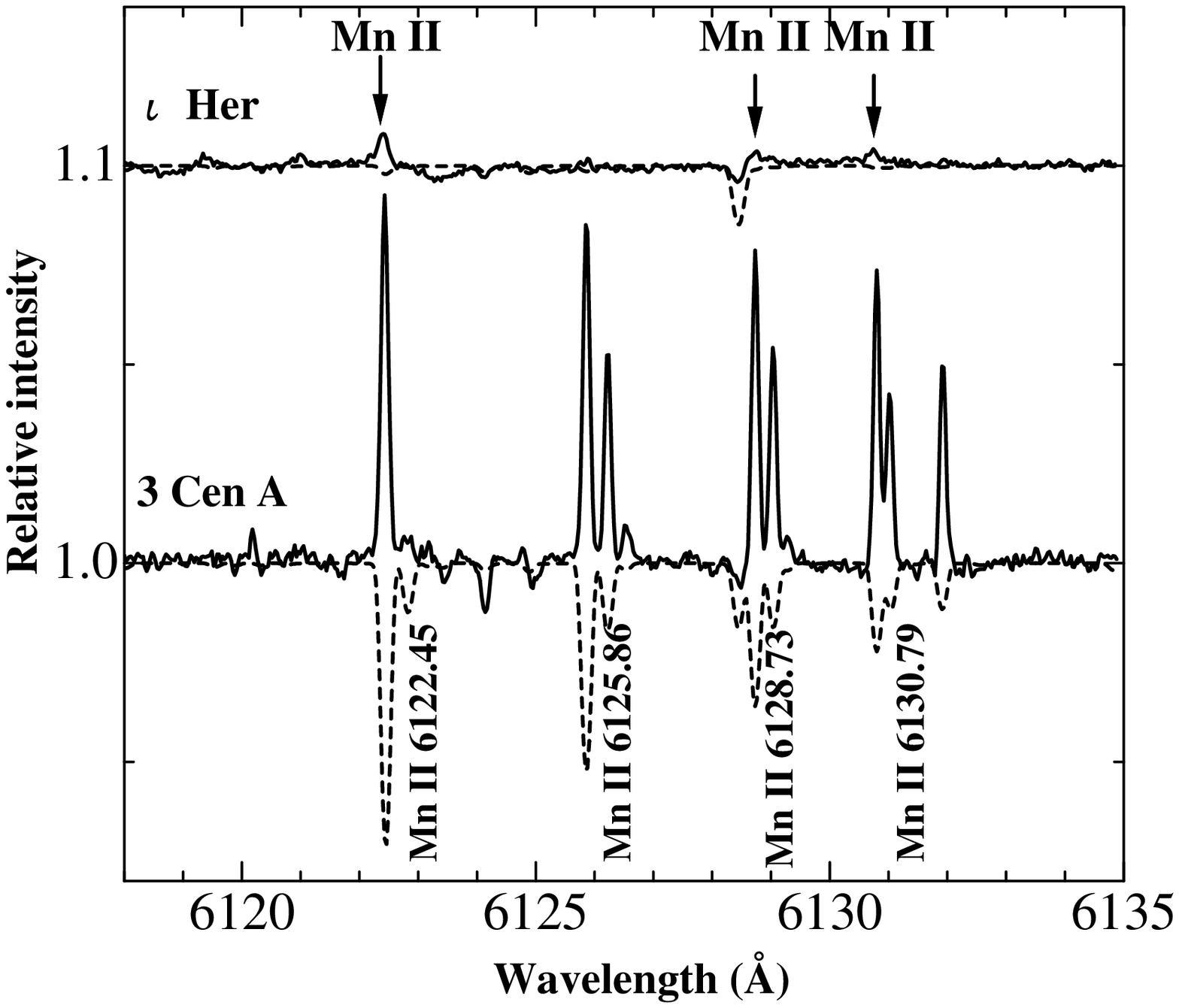}
       %%% \FigureFile(width,height){filename}
     \end{center}
     \caption{Emission lines of Mn~{\sc ii} in $\iota$  Her near 6120 \AA.
   Data of 3 Cen A are shown for comparison.  
Solid  and dashed lines show observed and simulated spectra, respectively. 
An over-abundance of Mn by 1.6 dex is assumed in the simulation  for
3 Cen A.  We notice three weak emission lines of 
Mn~{\sc ii} in $\iota$  Her (indicated by arrows).}
    \end{figure}
%--------------------------------------------------------
%**************************************

\section{The line list}

We pick up and register WELs observed in $\iota$ Her in the wavelength region from 
3900 \AA~ to 9980 \AA. We register an emission feature only when the feature 
is seen at the same position on both data observed on two nights  (2010 July  20 
and 2012 June  25).  In this way we can avoid possible contaminations from the
sky emission.
Measurements of central wavelength and peak intensity of a emission line 
have been done on two nights data. The emission equivalent width has been 
measured by directly integrating the profile over the local continuum. When
an emission line is located in the wing of a strong absorption line, we use 
the profile of the wing as the local continuum. 
In such a case, measured peak intensities sometimes become smaller than 1.0.
 Results of two measurements 
of a line are averaged and then registered. 

Line identifications have been done with the help of synthesized  spectrum of 
 $\iota$ Her using a program developed by \citet{takeda1995} under the assumption of LTE.
%*************** append**********
Atomic data given in the vald3 database  \citep{ryabchikova2015} are used in the present study.
%********************************
We adopt atmospheric parameters of $\iota$ Her ({\it $T_{\rm eff}$} = 17500 K 
and  log  $\it g$  = 3.80) from  \citet{nieva2012} and use an interpolated 
 ATLAS9 model atmosphere \citep{kurucz1993} in this study. 
Data of solar abundances given in \citet{asplund2009} are used in spectral 
simulations.

Figure 1 shows an example spectrum between 9390 \AA~ and 9415 \AA, where strong 
telluric absorption lines are overlapping on stellar spectrum.
The dashed line shows a synthesized spectrum. We expect three absorption
lines in this region on the simulation. They are N~{\sc i} at 9392.79 \AA, C~{\sc i} 9405.72 \AA,
and Si~{\sc ii}  9412.67 \AA~ and 9412.78 \AA~ (blended).  These three lines 
are clearly seen in emission in $\iota$ Her. The C~{\sc i} 9405.72 \AA~ line in this star 
has been analysed by \citet{alexeeva2016}.
A very weak emission feature is observed at 9406.7 \AA~ and this line
might be attributed to an Fe~{\sc ii} line at 9406.67 \AA. However, this line is not
registered because the line is too close to the neighboring  C~{\sc i} line.
Figure 2 shows another region near 9900 \AA. Spectra of three stars
($\gamma$ Peg,  $\iota$ Her, and 3 Cen A) 	are compared. Two clear
emission lines of Ca~{\sc ii} at 9890.63 \AA~ and C~{\sc ii} at 9903.46 \AA~
can be seen in both $\gamma$ Peg and  $\iota$ Her. 
The emission line of C~{\sc ii} at 9903.46 \AA~ has never been reported in 
stellar spectra.
The  Ca~{\sc ii}  line in $\iota$ Her has been analysed by \citet{sitnova2018a}. 
In the spectrum of 3 Cen A, emission lines of P~{\sc ii} and Mn~{\sc ii} 
are observed \citep{wahlgren2004}.

Tables 3 and 4 list 190 identified emission lines in  $\iota$ Her (hereafter 
referred to as table 3) . Observed wavelengths,
peak intensities, and emission equivalent widths together with relevant
physical quantities (laboratory wavelengths, log $\it gf$ values, potentials
 and configurations) are given.  Data of physical quantities have been adopted 
from the vald3 database.  
%*************** re-write from 14 to 15 01/17**************
We note that 15 emission lines have two or more candidates of 
identification (multiple entries). 
%***********************************

 There are 17 unidentified emission lines which can be seen on two data
of the star observed on different nights. We list measured wavelengths,
peak intensities, and emission equivalent widths of these unidentified lines in 
table 5. Consulting the line list of 3 Cen A prepared by \citet{wahlgren2004},
we find that two entries in table 5 have been given possible identifications
within $\pm$ 0.05 \AA~ of the observed line centers. For example,
the feature observed at 8731.72 \AA~ could  be identified as a Mn~{\sc ii} line at
8731.692 \AA, if we follow \citet{wahlgren2004}. We simulated 
the spectrum of $\iota$ Her around 8730 \AA~ using atomic data given 
%**********
the vald3 database 
%**********
and using the abundance of Mn in this star \citep{golriz2017} to find no 
significant line can be expected at 8731.7 \AA.  We find a similar
result for the line at 8838.33 \AA, too. Thus, we  leave these two lines in the 
table of unidentified lines.  We further notice that there are 13 common entries
of unidentified lines in table 5 and the list of emission lines in 3 Cen A 
prepared by \citet{wahlgren2004}. 

%************************************
%figure5
\begin{figure}
     \begin{center}
       \FigureFile(90mm,110mm){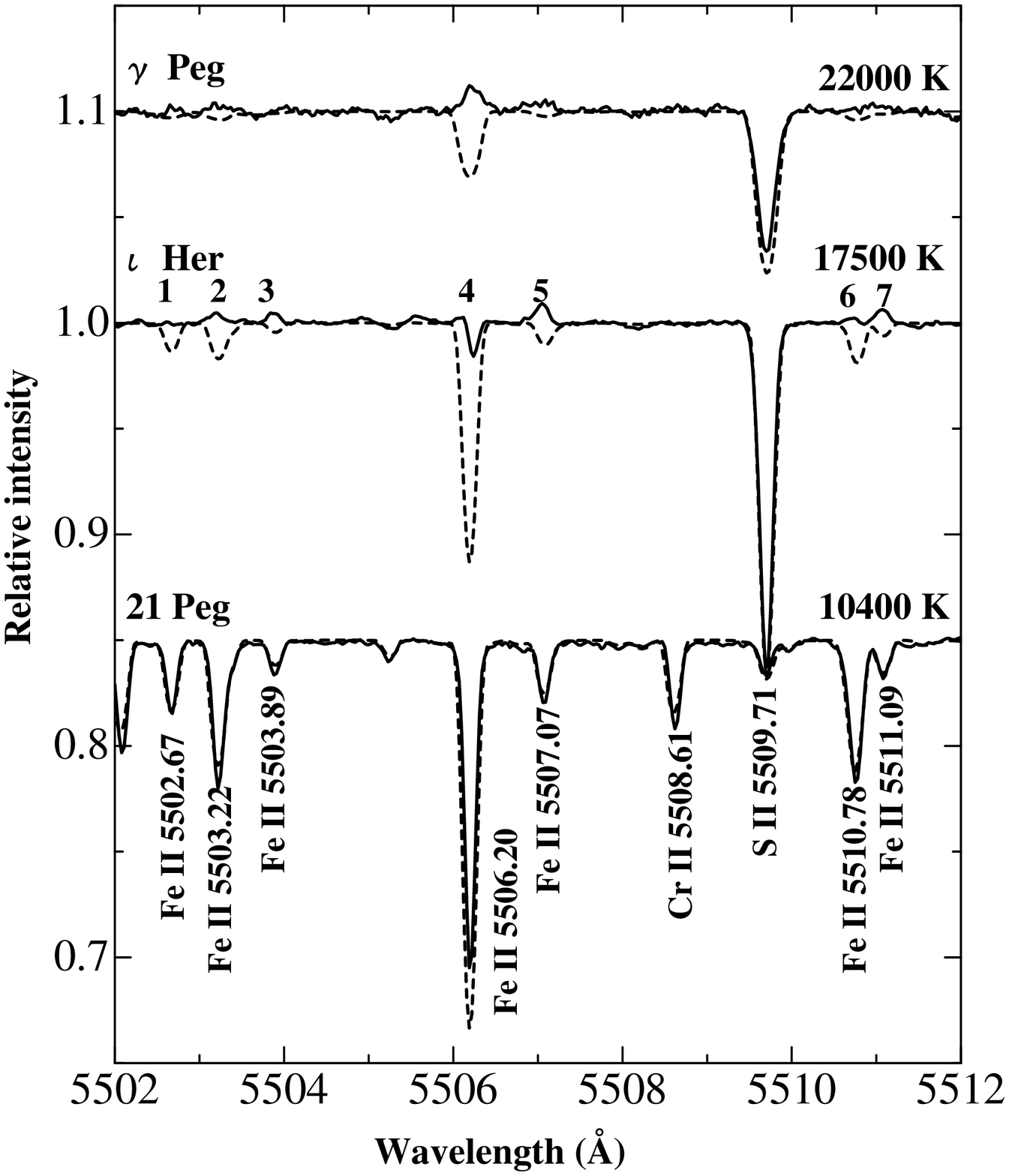}
       %%% \FigureFile(width,height){filename}
     \end{center}
     \caption{Weak emission lines of Fe~{\sc ii} near 5506 \AA.
       Observed (solid line) and simulated (dashed line) spectra 
       for three stars ($\gamma$ Peg,   $\iota$  Her and 21 Peg) are compared.
  %    Solid  and dashed lines show observed and simulated spectra, respectively.
  %      The solar abundances have been assumed in simulations.
}  
    \end{figure}
%----------------------------------------------------
%figure6
\begin{figure}
     \begin{center}
       \FigureFile(100mm,150mm){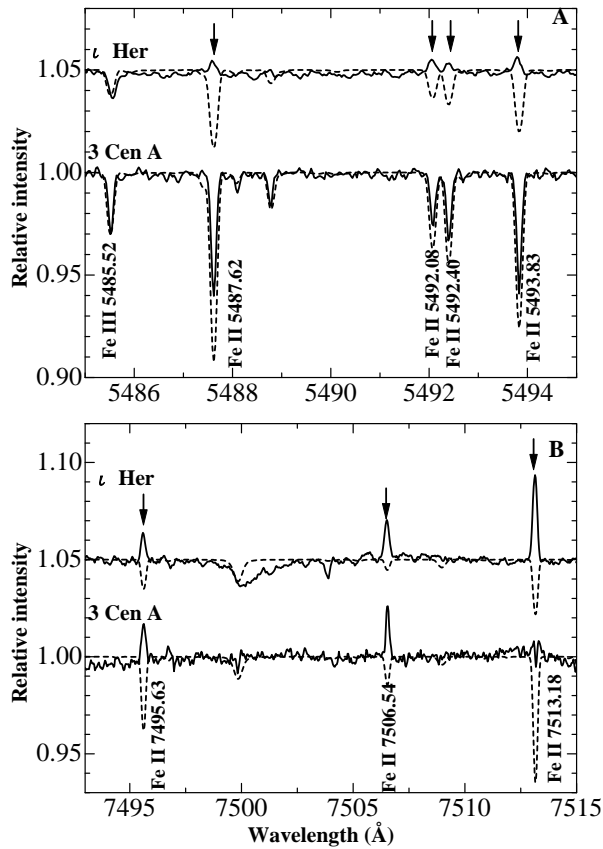}
       %%% \FigureFile(width,height){filename}
     \end{center}
\caption{Comparison of Fe~{\sc ii} lines between $\iota$  Her and 
3 Cen A in two wavelength regions. In the upper panel, Fe~{\sc ii} lines 
appear  as absorption in 3 Cen A, while they are in emission in the
lower panel.  
  Solid  and dashed lines show observed and simulated spectra, respectively.
Abundances of Fe obtained by \citet{sadakane2018} for these two stars have 
been used in simulations. }
  \end{figure} 
\setcounter {table}{2}
\begin{table*}
      \caption{Weak emission lines in $\iota$ Her: Element arrangement }\label{third}
%\fontsize{6.3pt}{5pt}\selectfont
\fontsize{7.5pt}{6pt}\selectfont
%\tiny
%\footnotesize
%\scriptsize
      \begin{center}
\begin{tabular}{cccccccccc}
\hline
\hline
Ion	&	$\lambda$(Obs)	& Peak Int.	&	E.W.	&	$\lambda$(Lab)	&	log $\it gf$ 	&	E${}_{low}$	&	E${}_{up}$	& \multicolumn{2}{c}{Config.} 		\\
      &    (\AA)     &           &    (m\AA) &   (\AA)  &                        &    cm${}^{-1}$      &   cm${}^{-1}$    &        &                     \\
\hline
C I	&	4932.08	&	1.002	&	0.6 	&	4932.049 	&	-1.658	&	61981.820 	&	82251.710 	&	2s2.2p.3s 1P*	&	2s2.2p.4p 1S	\\
C I	&	5052.16	&	1.006	&	1.4 	&	5052.144 	&	-1.303	&	61981.818 	&	81769.876 	&	2s2.2p.3s 1P*	&	2s2.2p.4p 1D	\\
C I	&	7115.19	&	1.002	&	1.0 	&	7115.170 	&	-0.934	&	69710.660 	&	83761.260 	&	2s2.2p.3p 3D	&	2s2.2p.4d 3F*	\\
C I	&	7116.99	&	1.002	&	0.4 	&	7116.981 	&	-0.907	&	69744.034 	&	83791.063 	&	2s2.2p.3p 3D	&	2s2.2p.4d 3F*	\\
C I	&	8335.16	&	1.044	&	17.2 	&	8335.147 	&	-0.437	&	61981.818 	&	73975.911 	&	2s2.2p.3s 1P*	&	2s2.2p.3p 1S	\\
C I	&	9061.46	&	1.027	&	10.1 	&	9061.430 	&	-0.347	&	60352.630 	&	71385.380 	&	2s2.2p.3s 3P*	&	2s2.2p.3p 3P	\\
\hline
        \end{tabular}
 \end{center}
\end{table*}
%************************************************************
%\newpage
%*********************************************
\setcounter {table}{2}
\begin{table*}
      \caption{continued}\label{fourh}
\fontsize{6.3pt}{5pt}\selectfont
%\fontsize{7.5pt}{6pt}\selectfont
%tiny
%\footnotesize
%\scriptsize
      \begin{center}
      \begin{tabular}{cccccccccc}
\hline
\hline
Ion	&	$\lambda$(Obs)	& Peak Int.	&	E.W.	&	$\lambda$(Lab)	&	log $\it gf$ 	&	E${}_{low}$	&	E${}_{up}$	& \multicolumn{2}{c}{Config.} 		\\
      &    (\AA)     &           &    (m\AA) &   (\AA)  &                        &    cm${}^{-1}$      &   cm${}^{-1}$    &        &                     \\
\hline
Fe II	&	5507.07 	&	1.009 	&	1.5 	&	5507.071 	&	-0.056 	&	84870.911 	&	103024.341 	&	3d6.(5D).4d f4D	&	3d6.(5D\textless4\textgreater).4f 2[2]*	\\
Fe II	&	5511.10 	&	1.007 	&	1.4 	&	5511.087 	&	-0.378 	&	85728.846 	&	103869.046 	&	3d6.(5D).4d 4S	&	3d6.(5D\textless1\textgreater).4f 2[2]*	\\
Fe II	&	5529.06 	&	1.012 	&	2.2 	&	5529.053 	&	-0.258 	&	84870.911 	&	102952.170 	&	3d6.(5D).4d f4D	&	3d6.(5D\textless4\textgreater).4f 2[3]*	\\
Fe II	&	5532.09 	&	1.005 	&	0.9 	&	5532.086 	&	-0.099 	&	84870.911 	&	102942.256 	&	3d6.(5D).4d f4D	&	3d6.(5D\textless4\textgreater).4f 2[3]*	\\
Fe II	&	5549.01 	&	1.008 	&	1.3 	&	5549.000 	&	-0.186 	&	84870.911 	&	102887.172 	&	3d6.(5D).4d f4D	&	3d6.(5D\textless4\textgreater).4f 2[4]*	\\
Fe II	&	5558.30 	&	1.005 	&	0.6 	&	5558.286 	&	-0.781 	&	85048.655 	&	103034.819 	&	3d6.(5D).4d f4D	&	3d6.(5D\textless4\textgreater).4f 2[2]*	\\
\hline
        \end{tabular}
 \end{center}
\end{table*}
%\newpage
\setcounter {table}{2}
\begin{table*}
      \caption{continued}\label{fifth}
%\fontsize{6.3pt}{5pt}\selectfont
\fontsize{7.5pt}{6pt}\selectfont
%\fontsize{7.3pt}{6pt}\selectfont
%tiny
%\footnotesize
%\scriptsize
      \begin{center}
      \begin{tabular}{cccccccccc}
\hline
\hline
Ion	&	$\lambda$(Obs)	& Peak Int.	&	E.W.	&	$\lambda$(Lab)	&	log $\it gf$ 	&	E${}_{low}$	&	E${}_{up}$	& \multicolumn{2}{c}{Config.} 		\\
      &    (\AA)     &           &    (m\AA) &   (\AA)  &                        &    cm${}^{-1}$      &   cm${}^{-1}$    &        &                     \\
\hline
Fe II	&	9159.29 	&	1.025 	&	6.0 	&	9159.271 	&	0.552 	&	99332.102 	&	110247.006 	&	3d6.(3H).5s e2H	&	3d6.(3H).5p 2I*	\\
Fe II	&	9166.32 	&	1.027 	&	7.3 	&	9166.302 	&	0.578 	&	103608.921	&	114515.453 	&	3d6.(3G).5s 2G	&	3d6.(3G).5p 2H*     \\
Fe II	&	9175.99 	&	1.027 	&	5.8 	&	9175.920 	&	0.479 	&	79885.523 	&	90780.623 	&	3d6.(5D).5s e4D	&	3d6.(5D).5p 4F*	\\
Fe II	&	9178.10 	&	1.047 	&	15.2 	&	9178.058 	&	0.362 	&	80178.005 	&	91070.566 	&	3d6.(5D).5s e4D	&	3d6.(5D).5p 4F*	\\
Fe II	&	9179.50 	&	1.035 	&	7.2 	&	9179.492 	&	0.128 	&	78237.709 	&	89128.569 	&	3d6.(5D).5s e6D	&	3d6.(5D).5p 6P*	\\
Fe II	&	9187.22 	&	1.041 	&	12.3 	&	9187.183 	&	0.242 	&	78237.709 	&	89119.452 	&	3d6.(5D).5s e6D	&	3d6.(5D).5p 6D*	\\
\hline
        \end{tabular}
 \end{center}
$\dagger$  Taken from the NIST database: \citet{kramida2018}.
\end{table*}
%************************************************************
\setcounter {table}{3}
\begin{table*}
      \caption{Weak emission lines in $\iota$ Her: Wavelength arrangement}\label{nineth}
%\fontsize{6.3pt}{5pt}\selectfont
\fontsize{7.5pt}{6pt}\selectfont
%\fontsize{7.3pt}{6pt}\selectfont
%\tiny
%\footnotesize
%\scriptsize
      \begin{center}
      \begin{tabular}{cccccccccc}
\hline
\hline
Ion	&	$\lambda$(Obs)	& Peak Int.	&	E.W.	&	$\lambda$(Lab)	&	log $\it gf$ 	&	E${}_{low}$	&	E${}_{up}$	& \multicolumn{2}{c}{Config.} 		\\
      &    (\AA)     &           &    (m\AA) &   (\AA)  &                        &    cm${}^{-1}$      &   cm${}^{-1}$    &        &                     \\
\hline
C I	&	4932.08 	&	1.002 	&	0.6 	&	4932.049 	&	-1.658 	&	61981.820 	&	82251.710 	&	2s2.2p.3s 1P*	&	2s2.2p.4p 1S	\\
Fe II	&	4948.78 	&	1.006 	&	0.8 	&	4948.791 	&	-0.031 	&	83459.721 	&	103661.038 	&	3d6.(5D).4d e6F	&	3d6.(5D\textless2\textgreater).4f 2[2]*	\\
C I	&	5052.16 	&	1.006 	&	1.4 	&	5052.144 	&	-1.303 	&	61981.818 	&	81769.876 	&	2s2.2p.3s 1P*	&	2s2.2p.4p 1D	\\
Fe II	&	5061.67 	&	1.005 	&	0.6 	&	5061.710 	&	0.284 	&	83136.508 	&	102887.172 	&	3d6.(5D).4d e6F	&	3d6.(5D\textless4\textgreater).4f 2[4]*	\\
Fe II	&	5067.90 	&	1.006 	&	1.4 	&	5067.890 	&	-0.078 	&	83308.241 	&	103034.819 	&	3d6.(5D).4d e6F	&	3d6.(5D\textless4\textgreater).4f 2[2]*	\\
Fe II	&	5075.72 	&	1.005 	&	1.0 	&	5075.760 	&	0.184 	&	84326.967 	&	104022.961 	&	3d6.(5D).4d 6P	&	3d6.(5D\textless0\textgreater).4f 2[3]*	\\
\hline
        \end{tabular}
 \end{center}
\end{table*}
\setcounter {table}{3}
\begin{table*}
      \caption{continued}\label{tenth}
%\fontsize{6.3pt}{5pt}\selectfont
\fontsize{7.5pt}{6pt}\selectfont
%tiny
%\footnotesize
%\scriptsize
      \begin{center}
      \begin{tabular}{cccccccccc}
\hline
\hline
Ion	&	$\lambda$(Obs)	& Peak Int.	&	E.W.	&	$\lambda$(Lab)	&	log $\it gf$ 	&	E${}_{low}$	&	E${}_{up}$	& \multicolumn{2}{c}{Config.} 		\\
      &    (\AA)     &           &    (m\AA) &   (\AA)  &                        &    cm${}^{-1}$      &   cm${}^{-1}$    &        &                     \\
\hline
Fe II	&	7574.05 	&	1.011 	&	2.2 	&	7574.024 	&	-0.074 	&	98445.410 	&	111644.798 	&	3d6.(3H).5s e4H	&	3d5.(4G)4s.4p(1P*)	\\
Fe II	&	7579.43 	&	1.099 	&	1.8 	&	7579.322 	&	-1.063 	&	78725.790 	&	91915.950 	&	3d6.(5D).5s e6D	&	3d5.(4D).4s.4p.(3P) 6F*	\\
          &	           	&		&		&	7579.559 	&	-1.132 	&	78725.822 	&	91915.570 	&	3d6.(5D).5s e6D	&	3d5.(4D).4s.4p.(3P*) 6F*	\\
Fe II	&	7592.70 	&	1.006 	&	1.7 	&	7592.685 	&	-0.095 	&	98568.907 	&	111735.854 	&	3d6.(3H).5s e4H	&	3d6.(3F).5p 4H*	\\
Fe II	&	7731.72 	&	1.016 	&	3.9 	&	7731.688 	&	-0.108 	&	78237.709 	&	91167.936 	&	3d6.(5D).5s e6D	&	3d6.(5D).5p w6P*	\\
Fe II	&	7755.59 	&	1.015 	&	3.3 	&	7755.571 	&	-0.587 	&	79439.496 	&	92329.905 	&	3d6.(5D).5s e4D	&	3d5.(4G).4s.4p.(3P*) v4F*	\\
\hline
 \end{tabular}
 \end{center}
\end{table*}

\setcounter {table}{3}
\begin{table*}
      \caption{continued}\label{eleventh}
%\fontsize{6.3pt}{5pt}\selectfont
\fontsize{7.5pt}{6pt}\selectfont
%tiny
%\footnotesize
%\scriptsize
      \begin{center}
      \begin{tabular}{cccccccccc}
\hline
\hline
Ion	&	$\lambda$(Obs)	& Peak Int.	&	E.W.	&	$\lambda$(Lab)	&	log $\it gf$ 	&	E${}_{low}$	&	E${}_{up}$	& \multicolumn{2}{c}{Config.} 		\\
      &    (\AA)     &           &    (m\AA) &   (\AA)  &                        &    cm${}^{-1}$      &   cm${}^{-1}$    &        &                     \\
\hline
Fe II	&	9070.49 	&	1.007 	&	1.9 	&	9070.462 	&	-0.377 	&	80178.005 	&	91199.776 	&	3d6.(5D).5s e4D	&	3d6.(5D).5p 4D*	\\
Fe II	&	9075.58 	&	1.042 	&	12.9 	&	9075.538 	&	0.151 	&	79885.523 	&	90901.130 	&	3d6.(5D).5s e4D	&	3d6.(5D).5p 4P*	\\
Fe II	&	9077.40 	&	1.032 	&	10.2 	&	9077.394 	&	0.069 	&	79885.523 	&	90898.878 	&	3d6.(5D).5s e4D	&	3d6.(3P1).4p 4P*	\\
C I	&	9078.30 	&	1.034 	&	13.5 	&	9078.280 	&	-0.581 	&	60352.630 	&	71364.900 	&	 2s2.2p.3s 3P*	&	2s2.2p.3p 3P	\\
C I	&	9088.55 	&	1.050 	&	13.0 	&	9088.509 	&	-0.430 	&	60352.639 	&	71352.525 	&	2s2.2p.3s 3P*	&	2s2.2p.3p 3P	\\
C I	&	9094.88 	&	1.088 	&	29.2 	&	9094.829 	&	0.151 	&	60393.148 	&	71385.390 	&	2s2.2p.3s 3P*	&	2s2.2p.3p 3P	\\
\hline 
\end{tabular}
 \end{center}
$\dagger$  Taken from the NIST database: \citet{kramida2018}.
\end{table*}

%*************************************************************
%*************************************

\section{Discussion}

Our survey of weak emission lines (WELs) in $\iota$ Her resulted in a  list of 
190 identified lines of 10 ions. We compare in table 6 numbers of detected
emission lines in $\iota$ Her and in 3 Cen A. Relative abundances of elements
with respect to the Sun in both stars are shown in the table.
3 Cen A has nearly the same atmospheric parameters  ({\it $T_{\rm eff}$} = 
17500 K and  log  $\it g$  = 3.8, \cite{castelli1997}) as those of $\iota$ Her. 

We detect in $\iota$ Her emission lines of six ions (C~{\sc ii}, N~{\sc i}, 
 Cr~{\sc ii}, Mn~{\sc ii},  and Ni~{\sc ii}), which have not been 
reported earlier in this star. 
The C~{\sc ii} emission line at 9903.46 \AA~ has not been reported in any
star. As illustrated in figure 2, the C~{\sc ii} emission line is strong in a
B2 IV star $\gamma$ Peg. The emission line is weak but
definitely present in $\iota$ Her.
Although emission lines of N~{\sc i} have been reported in a few chemically peculiar
stars such as 3 Cen A \citep{wahlgren2004}, our detections of three
emission lines of N~{\sc i} is the first case among normal stars (figure 1).  
We find several clear emission lines of Si~{\sc ii} in $\iota$ Her. Two
emission lines of  Si~{\sc ii} at 5688.82 \AA~ and at 6239.61\AA~ reported in 
\citet{sadakane2017} are weak. The newly recorded pair of Si~{\sc ii}
emission lines at 7848.82 \AA~ and at 7849.72 \AA~ (figure 3) are strong and 
undisturbed by other lines. These two lines 
are good candidates to be used in quantitative analyses.  

Emission lines of Mn~{\sc ii} between 6120 \AA~ and 6135 \AA~ have been
reported in many CP stars (e.g., \cite{sigut2000}, \cite{wahlgren2000}), but
have not been detected in normal stars. Figure 4 show a comparison of spectra
of  $\iota$ Her and 3 Cen A in the region between 6118 \AA~ and 6135 \AA.
We can see three weak emission features (indicated by arrows) in  $\iota$ Her
just at the positions of three Mn~{\sc ii} lines at 6122.45 \AA, 6128.73 \AA, and
at 6130.79 \AA. Because these emission features are visible on data of
two nights (observed on 2010 July 20 and 2012 June 25) and the SN ratios
are higher than $\sim$ 750 in both data, we conclude positive detections 
of three emission lines of Mn~{\sc ii} in this star.

Over 130 emission lines of Fe~{\sc ii} have been listed in table 3. They  all
originate from highly excited states of Fe~{\sc ii}.
We notice an interesting feature when comparing  emission lines of Fe~{\sc ii} 
in $\iota$ Her with those observed in a hotter star $\gamma$ Peg ({\it $T_{\rm eff}$}
= 22000 K, \cite{nieva2012}). 
Figure 5 compares Fe~{\sc ii} features between 5502 \AA~ and 5512 \AA.
There are  relatively strong seven lines  of  Fe~{\sc ii}  labelled as 1 to 7, 
all of which arise from highly excited  states (near 10.5 eV), in this region. 
 Four lines labelled as 2, 3, 5, and   7 have been registered as
 emission lines in table 3.  Peak intensities of the remaining three lines in  $\iota$  Her 
are too weak to be recognized as emission lines. We notice that  line 4 at 5506.20 \AA~ 
(the  line has the largest log $\it gf$ value among the seven lines) shows no emission
component but looks like nearly filled-in. The line  appears as a 
clear emission line in a hotter  star $\gamma$ Peg. 
All the other six lines also appear as weaker  emission lines in $\gamma$ Peg. 
On the other hand, these seven Fe~{\sc ii}  lines appear as absorption
 lines in a cooler star 21 Peg ({\it $T_{\rm eff}$} = 10400 K, \cite{fossati2009}). 
Equivalent widths of these seven absorption lines of Fe~{\sc ii} 
are in accordance with the LTE  simulation for 21 Peg.
We guess that   $\iota$  Her might be a transitional case.  When we go from
cooler to hotter stars, weak Fe~{\sc ii}  lines turn into emission at lower temperature
 and  strong lines remain as weak  absorption lines in transitional cases. Then 
strong lines turn into emission at even higher temperature.

In table 6,  we notice a correlation between the number of detected emission lines 
and the abundances. In the case of C~{\sc i}, we detected 13 emission lines in 
 $\iota$ Her, while only one emission line of this ion has been reported in 
3 Cen A by \citet{wahlgren2004}. Carbon has a normal abundance in $\iota$ Her,
while the element is under-abundant in 3 Cen A \citep{sadakane2018}. 
This relation can be found in the case of Al~{\sc ii}, too.
 We detect three emission line of Al~{\sc ii} in  $\iota$ Her,
in which Al is normal. \citet{wahlgren2004} listed no  emission line of   Al~{\sc ii}  in 
3 Cen A, in which Al is definitely under-abundant  \citep{sadakane2018}. 
\citet{castelli1997} showed that  P, Mn, and Cu are all over-abundant in 3 Cen A and  
\citet{wahlgren2004} reported many emission line of P~{\sc ii} , Mn~{\sc ii}, and  Cu~{\sc ii}
in this star. 
We detected no emission line of  P~{\sc ii} and  Cu~{\sc ii} and found only three
very weak emission lines of Mn~{\sc ii} in $\iota$ Her. 
Similar relations can be found for Ti~{\sc ii} and Ni~{\sc ii}, too. These findings support
the suggestion as to the dependence of strengths of emission lines on the element
abundance \citep{wahlgren2000}.

%**********************************
%\newpage
\setcounter {table} {4}\begin{table}
      \caption{Unidentified emission lines} \label{seventh}
%\tiny
%\footnotesize
\scriptsize
      \begin{center}
      \begin{tabular}{cccc}
\hline\hline
$\lambda$ & Peak intensity & E. W.   & note$^{*}$      \\
(\AA)      &                & (m\AA)      &        \\
\hline
6476.60 	&	1.005 	&	0.6  &  	\\	
8129.37 	&	1.010 	&	3.1  & unknown	\\	
8215.72 	&	1.014 	&	2.6  & 	          \\	
8304.26 	&	1.012 	&	2.0  & unknown	\\	
8459.34 	&	1.000 	&	2.6  & unknown	\\	
8682.87 	&	0.977 	&	3.1  & unknown	\\	
8719.69 	&	1.000 	&	1.3  & unknown	\\	
8731.72 	&	0.960 	&	1.8  & Mn~{\sc ii}  8731.692 \\	
8767.58 	&	0.968 	&	5.0  & unknown	\\	
8780.16 	&	0.982 	&	2.8  & unknown	\\	
8788.38 	&	0.993 	&	1.2  & unknown	\\	
8825.57 	&	1.017 	&	5.4  & unknown 	\\	
8838.33 	&	0.999 	&	5.7  & Ti~{\sc ii}  8838.415 	\\	
8882.14 	&	1.000 	&	3.2  & 	\\	
8910.43 	&	1.020 	&	8.0  & unknown	\\	
8953.90 	&	1.013 	&	5.1  & unknown	\\	
8961.82 	&	1.016 	&	3.9  & 	\\	
\hline
 \end{tabular}
 \end{center}
* Entries given in the line list of 3 Cen A \citep{wahlgren2004}. 
\end{table}

%\newpage

\setcounter {table}{5}\begin{table}
\caption{Number of registered emission lines} \label{eightth}
%\tiny
\footnotesize
%\scriptsize
      \begin{center}
      \begin{tabular}{lcccc}
\hline\hline
Ion & $\iota$ Her &  & 3 Cen A & \\
     &     N            & [X/H] & N  &  [X/H]  \\
\hline
 C~{\sc i}    & 13 & -0.05$^{a}$   & 1  & -0.75$^{a}$    \\
 C~{\sc ii}   & 1   & ...        & 0  & ...         \\
 N~{\sc i}     & 3   & 0.03$^{a}$    & 3   & 0.44$^{a}$     \\
 N~{\sc ii}    & 0   & ...        & 1   & ...        \\
 Al~{\sc ii}   & 3    & 0.13$^{a}$    & 0   &  -1.37$^{a}$  \\
 Si~{\sc ii}   & 12  & -0.01$^{a}$   & 6  &  -0.15$^{a}$    \\
 P~{\sc ii}    & 0   & 0.04$^{a}$        & 44  & 2.01$^{a}$     \\
 Ca~{\sc ii}   & 8   &-0.02$^{b}$    & 3   & -0.14$^{c}$        \\
 Ti~{\sc ii}   & 0    & -0.16$^{a}$   & 2   &  0.32$^{a}$   \\
 Cr~{\sc ii}   & 2   & -0.07$^{a}$   & 4    &  0.08$^{a}$  \\
 Mn~{\sc ii}   & 3  &-0.2$^{b}$        & 34   &  1.60$^{a}$  \\
%************* updated **********
 Fe~{\sc ii}   & 135 & -0.37$^{a}$   & 137$^{e}$ & 0.42$^{a}$   \\
%*******************************
 Co~{\sc ii}   & 0    & -0.48$^{b}$      & 2  & ...         \\
 Ni~{\sc ii}   & 10   &   -0.27$^{a}$  & 30 & 0.46$^{a}$     \\
 Cu~{\sc ii}   & 0  & -1.17$^{b}$         & 15 & 2.00$^{d}$    \\
 Hg~{\sc ii}   & 0  & 1.95$^{b}$         & 1   & 3.30$^{a}$     \\
\hline
\end{tabular}
 \end{center}
 a: Sadakane and Nishimura (2018), b: Golriz and Landstreet (2017) 
 c: Hardorp et al. (1968), d: Castelli et al. (1997), e: nine additional
Fe~{\sc ii} lines have been detected in 3 Cen A among emission lines
noted as unknown in \citet{wahlgren2004}.

\end{table}  
%------------------------------------------------------
%figure7
\begin{figure}
     \begin{center}
       \FigureFile(90mm,110mm){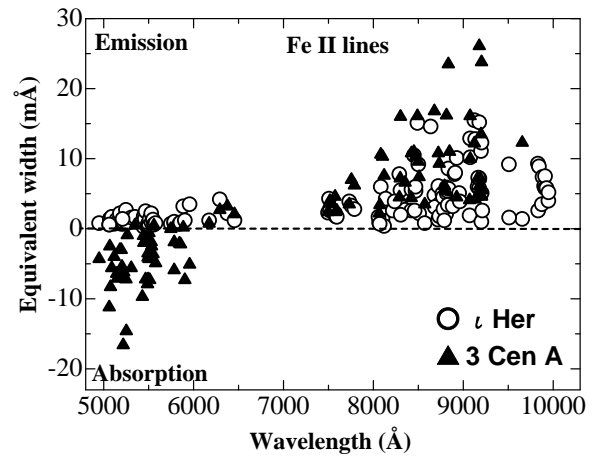}
       %%% \FigureFile(width,height){filename}
     \end{center}
     \caption{Equivalent widths of Fe~{\sc ii} lines in $\iota$  Her (open
     circles) and in 3 Cen A (filled triangles) plotted against the wavelength.
     Data of $\iota$  Her are taken from table 3. Equivalent widths of
     corresponding lines in 3 Cen A in the shorter  wavelength side of  6000 \AA~ 
     are measured using our own data, while those in  the longer  wavelength side 
     are  taken from \citet{wahlgren2004}. Emission and absorption lines are plotted
     in the upper (positive) and lower (negative) parts, respectively.}
  \end{figure}
%*******************

We list 46 emission  lines of Fe~{\sc ii} in the shorter wavelength side of  6000 \AA~ in 
 $\iota$  Her.
Curiously, however,  most of  these emission lines  are missing in the list of
3 Cen A.  
\citet{wahlgren2004} measured  only one emission  line of Fe~{\sc ii} 
at 5794.90 \AA~ between 4900 \AA~ and 6000 \AA.   The line is not observed in $\iota$ Her.
They listed the line as an Fe~{\sc ii} 
line at 5794.726 \AA~ or a Si~{\sc ii} line at 5794.890 \AA. 
 A weak emission line can be seen at 5794.89 \AA~ on our data
of 3 Cen A and  we suggest  the Si~{\sc ii} line at 5794.890 \AA~ as a better
candidate because of the coincidence between the observed and laboratory wavelengths.
Close inspections 
of our spectral data of 3 Cen A reveal that most of these apparently missing  
Fe~{\sc ii} lines appear as absorption lines in this star (figure 6, panel A). 
It is natural that these lines are not included in the list of emission lines.
On the contrary, most of Fe~{\sc ii} WELs found 
in $\iota$  Her also appear as emission lines in 3 Cen A in the longer wavelength side of  6000 \AA~(figure 6, panel B). 
Measured  equivalent widths of Fe~{\sc ii} emission lines in $\iota$  Her (taken from table 3) 
are plotted against the wavelength in figure 7. Those of corresponding lines 
in 3 Cen A are shown for comparison. We can see an upward trend (from left to right) in
equivalent widths of Fe~{\sc ii} lines of 3 Cen A. This shows a transition from 
absorptions to emissions occurs at around 6000 \AA.  
The contrast  in the behaviours of Fe~{\sc ii} lines in the shorter
 wavelength side of  6000 \AA~ 
between $\iota$  Her and 3 Cen A might be related to a phenomenon 
reported in \citet{wahlgren2000}. They  found an anti-correlation between the
strength of emission line of Mn~{\sc ii} at 6122.43 \AA~ and the Mn abundance
enhancement. Among HgMn stars they observed, stars with large values of
[Mn/H] (higher than $\sim$ 1.3 dex) do not show emission in this line.
The abundance of Fe in 3 Cen A is higher than that in   $\iota$  Her 
by at least 0.3 dex \citep{sadakane2018}, a similar mechanism may be involved. 
Because these two stars have nearly the same atmospheric parameters 
 ({\it $T_{\rm eff}$} and log $\it g$), a physical interpretation  independent on
these two parameters is needed.
  
\vskip 3mm

This research is based on  CFHT public science archive data retrieved through the Canadian
Astronomy Data Centre and has made use of the SIMBAD database, operated
by CDS, Strasbourg, France. 
%**************** append ***********************
We thank the referee for providing valuable comments which helped to improve
the manuscript.
%*******************

\end{document}